\documentclass[pra,aps,twocolumn,superscriptaddress]{revtex4}
\usepackage{amsmath,amsfonts}
\usepackage{latexsym}
\usepackage{braket}
\usepackage{amssymb}
\usepackage[compact]{titlesec}
\usepackage{tabularx}
\usepackage[normalem]{ulem}
\usepackage{float}
\usepackage{graphicx}
\usepackage{color}
\usepackage[noxcolor]{pstricks}
\usepackage{pst-all}
\usepackage[color]{changebar}
\usepackage[caption=false]{subfig}
\usepackage{lipsum}

\newcommand{\be}{\begin{equation}}
\newcommand{\ee}{\end{equation}}
\newcommand{\eins}{\mbox{$1 \hspace{-1.0mm}  {\bf l}$}}
\newcommand{\ba}{\begin{aligned}}
\newcommand{\ea}{\end{aligned}}

\def\squareforqed{\hbox{\rlap{$\sqcap$}$\sqcup$}}
\def\qed{\ifmmode\squareforqed\else{\unskip\nobreak\hfil
\penalty50\hskip1em\null\nobreak\hfil\squareforqed
\parfillskip=0pt\finalhyphendemerits=0\endgraf}\fi}
\def\endenv{\ifmmode\;\else{\unskip\nobreak\hfil
\penalty50\hskip1em\null\nobreak\hfil\;
\parfillskip=0pt\finalhyphendemerits=0\endgraf}\fi}

\begin{document}

\title{Complementarity between entanglement-assisted and quantum distributed random access code}

\author{Alley Hameedi}
\affiliation{Department of Physics, Stockholm University, S-10691 Stockholm, Sweden}

\author{Debashis Saha}
\affiliation{Institute of Theoretical Physics and Astrophysics, University of Gda\'{n}sk, 80-952 Gda\'{n}sk, Poland}

\author{Piotr Mironowicz}
\affiliation{Department of Algorithms and System Modeling, Faculty of Electronics, Telecommunications and Informatics, Gda\'{n}sk University of Technology, Gda\'{n}sk 80-233, Poland}
\affiliation{Institute of Theoretical Physics and Astrophysics, University of Gda\'{n}sk, 80-952 Gda\'{n}sk, Poland}

\author{Marcin Paw\l{}owski}
\affiliation{Institute of Theoretical Physics and Astrophysics, University of Gda\'{n}sk, 80-952 Gda\'{n}sk, Poland}

\author{Mohamed Bourennane}
\affiliation{Department of Physics, Stockholm University, S-10691 Stockholm, Sweden}



\begin{abstract}
Collaborative communication tasks such as random access codes (RACs) employing quantum resources have
manifested great potential in enhancing information processing capabilities beyond the classical limitations. The
two quantum variants of RACs, namely, quantum random access code (QRAC) and the entanglement-assisted
random access code (EARAC), have demonstrated equal prowess for a number of tasks. However, there do exist
specific cases where one outperforms the other. In this article, we study a family of $3 \rightarrow 1$ distributed RACs \cite{network} and present its general construction
of both the QRAC and the EARAC. We demonstrate that, depending on the function of inputs that is sought, if
QRAC achieves the maximal success probability then EARAC fails to do so and vice versa.Moreover, a tripartite
Bell-type inequality associated with the EARAC variants reveals the genuine multipartite nonlocality exhibited
by our protocol. We conclude with an experimental realization of the $3 \rightarrow 1$ distributed QRAC that achieves
higher success probabilities than the maximum possible with EARACs for a number of tasks. 

\end{abstract}

\maketitle

\section{Introduction}
Quantum theory has revolutionized the field of information
processing in the last few decades. The advantages offered by
quantum resources can be exploited in two distinct ways. The
first one involves spatially correlated shared entangled states
followed by classical communication of the quantum measurement outcome performed on these states,whereas in the second
scenario, a prepared quantum system is communicated that can
be later measured to extract information. Teleportation \cite{tele}, remote state preparation \cite{rsp}, nonlocal games \cite{chsh,cc}, quantum key distribution \cite{bb92} are a few applications that use the resources of the first type. In the prepare and measure scenario, several information processing protocols can also be realized, for example: quantum key distribution \cite{bb84,qkd}, randomness certification \cite{rc}, characterization of quantum correlations \cite{ic}, dimension witness \cite{dw}, parity oblivious multiplexing \cite{pom} have been proposed. Many of these protocols have also been experimentally implemented \cite{dwexp,rac-d,2s2r} and random access codes (RACs)\cite{rac} have been a powerhouse fueling the most of them.
In RAC, the preparation device encodes a bit-string into a single bit before communicating it to the measurement device whose task is to retrieve one of the arbitrarily chosen bits from the string. Quantum resources are used by either sending a quantum system through a quantum channel; the protocol is then called simply a quantum random access code (QRAC), or by sharing quantum entanglement among the devices with classical communication channel; the name entanglement assisted random access code (EARAC) is then used. Here we consider a generalized version of RAC in which the measurement device is asked to retrieve a particular function of the preparation device's inputs. The set of functions that can be given to the measurement device is called a task.

Although these two manifestations of quantum resources are equally efficient in many cases, this is not always the case. Their nonequivalence has been shown \cite{earac,magical7,spvssq} by considering distinct scenarios. In the case of a unit channel capacity, EARACs perform better than QRACs with classical shared randomness. Contrarily, for higher dimensional system, QRAC outperform EARAC in a few particular applications. Here we demonstrate that, remarkably, for the same application, either EARAC or QRAC is better, depending on the task. In other words, with respect to the optimal implementation of quantum resources, EARAC and QRAC are complementary to each other.

We begin by considering the $3 \rightarrow 1$ distributed RAC scenario, introduced in \cite{network}. This is the simplest form of a communication network that consists of three components - preparation, transformation and measurement devices. We then extensively study this variant of RAC using the two different types of quantum resources mentioned earlier. First, we present the EARAC protocol with Greenberger-Horne-Zeilinger (GHZ) states that leads to the maximal success probability for some task. In addition, we also propose a tripartite Bell-type inequality associated with the EARAC approach, which reveals that the protocol manifests genuine multipartite nonlocality. In the next section, we will demonstrate the general construction of the QRAC protocol that leads to the maximal success probability for a different task. Further, it is shown that, depending on the task, when QRAC has the maximal success probability, the EARAC fails and vice versa. Our results not only signify the quantum advantage over classical but more interestingly also point out the versatility of these quantum resources. Finally, we illustrate the experimental realization of the $3 \rightarrow 1$ distributed QRAC for a number of tasks.   \\

\section{Distributed random access code}
\begin{figure}[!htb]
\centering
\includegraphics[scale=0.5]{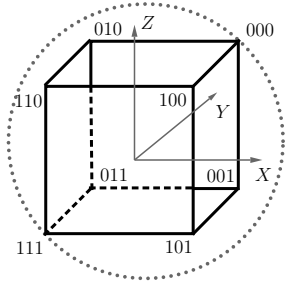}
\caption{The encoding quantum states given by Eq. \eqref{qs} in Bloch sphere for input $x_0x_1x_2$ for the standard $3 \rightarrow 1$ RAC.}
\label{fig2}
\end{figure}

 The standard $3 \rightarrow 1$ RAC is a communication complexity problem defined in prepare and measure scenario. The preparation device receives a string of three bits $x=(x_0,x_1,x_2)$ randomly chosen from a uniform distribution and communicates a two dimensional system to the measurement device. The measurement device also receives an input $y \in \{0,1,2\}$ with an aim to guess $x_y$. The average success probability (see Eq.\eqref{avgsp}) of guessing $x_y$ is $\frac{3}{4}$ for classical system. While QRAC and EARAC offer the same success probability, say $P^Q = \frac{1+\sqrt{3}}{2\sqrt3}\approx 0.7887$ \cite{rac,earac}.
In an optimal quantum strategy, the encoding quantum states are given by,
\begin{equation}\begin{split}
& |\psi(\theta,\phi)\rangle = \cos(\theta) |0\rangle + e^{i \phi} \sin(\theta) |1\rangle, \\
& \theta = \cos^{-1}\left( \sqrt{\frac{\sqrt{3}+ (- 1)^{x_2}}{2\sqrt3}} \right),\\
& \phi = \frac{\pi}{4}\left(1+4x_0+2(x_0\oplus x_1)\right).
\end{split}
\label{qs}
\end{equation}
These states are the eight vertices of the cube fit inside the qubit representation of Bloch sphere as shown in Fig. \ref{fig2}.
The decoding strategy is to measure the quantum state in the bases $\{\ket{\circlearrowright}, \ket{\circlearrowleft}\}, \{\ket{+},\ket{-}\}$ and $\{\ket{0},\ket{1}\}$ for output $z \in \{0,1\}$, in which the basis choice corresponds to the input $y=0,1,2$ respectively. These are the mutually unbiased bases $\sigma_Y,\sigma_X,\sigma_Z$, where $\ket{\circlearrowleft} = \frac{1}{\sqrt{2}} (\ket{0} + i \ket{1}),\ket{\circlearrowright} = \frac{1}{\sqrt{2}} (\ket{0} - i \ket{1}),\ket{+} = \frac{1}{\sqrt{2}} (\ket{0} + \ket{1}),\ket{-} = \frac{1}{\sqrt{2}} (\ket{0} - \ket{1}).$

In the distributed version of such a communication task (see Figure \ref{fig}), the preparation device is split into two devices such that the first device receives two bits $x_0,x_1$ and the second device receives only $x_2$. Besides, the communication channel capacity between these two devices is restricted to be one. However, all the devices are allowed to share classical randomness. Here, we consider a general task where the output $z$ is a function of $x,y$, and the figure of merit is defined by the average success probability,
\begin{equation} \label{avgsp}
P = \frac{1}{24} \sum_{x,y} p(z=f(x,y)|x,y).
\end{equation}
By considering all classical deterministic strategies, it can be checked that when $f(x,0),f(x,1),f(x,2)$ are independent of each other (i.e., one of them has zero information about another), then the upper bound of $P$ using classical channel is $\frac{2}{3}$. This value can be achieved by communicating the majority bit of two input bits among three to the measurement device.
Although the value of $P$ decreases from $\frac{3}{4}$ to $\frac{2}{3}$ in the classical distributed RAC, there exist quantum strategies that lead to the maximal success probability for several forms of $f(x,y)$. Since distribution of inputs in two devices imposes additional constraints over the standard one, $P^Q$ is the upper bound for the quantum success probability of all the tasks we consider in the distributed RAC.  \\

\section{Entanglement assisted distributed random access code}
\begin{figure}[htp]

\subfloat[Distributed RAC with a classical channel]{%
  \includegraphics[scale=0.55]{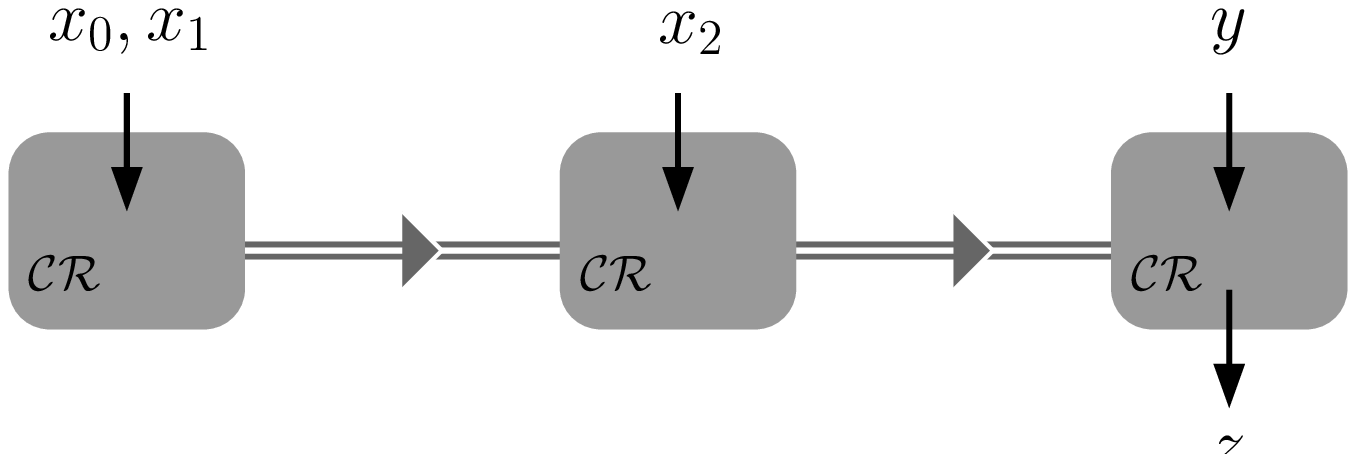}%
}
\\

\subfloat[Distributed RAC with shared entanglement and a classical channel]{%
  \includegraphics[scale=0.55]{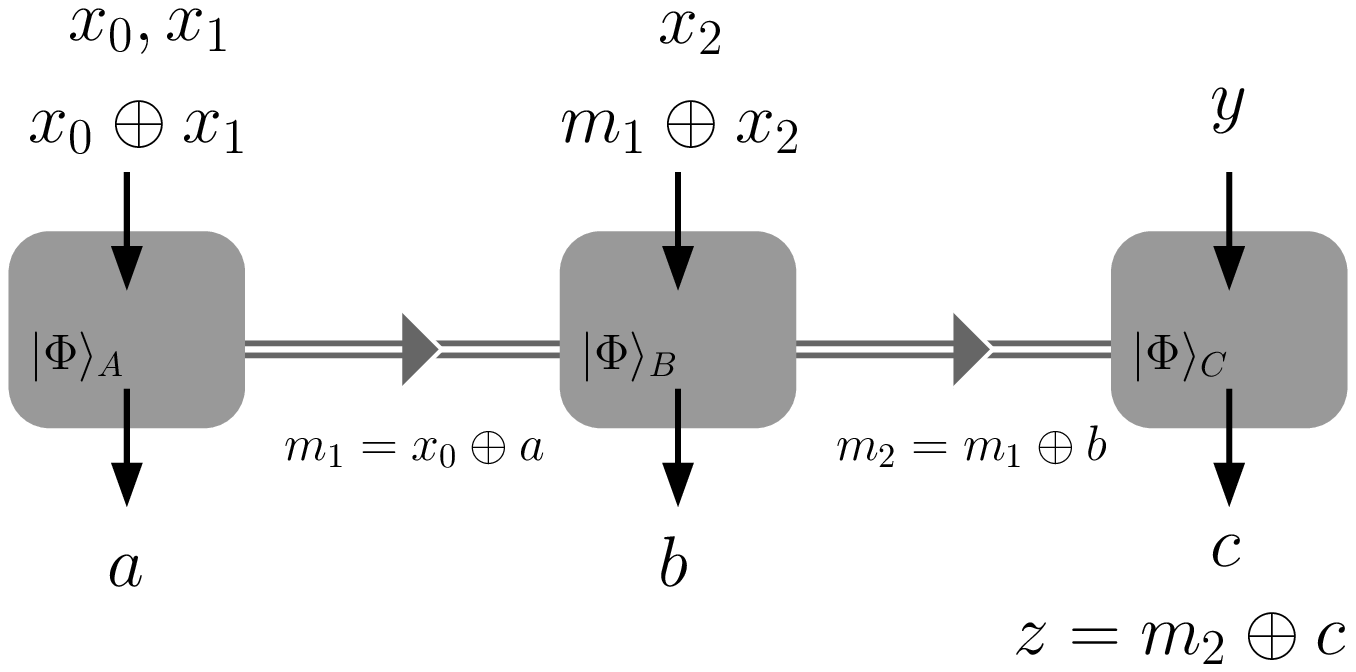}%
}
\\

\subfloat[Distributed RAC with a quantum channel]{%
  \includegraphics[scale=0.55]{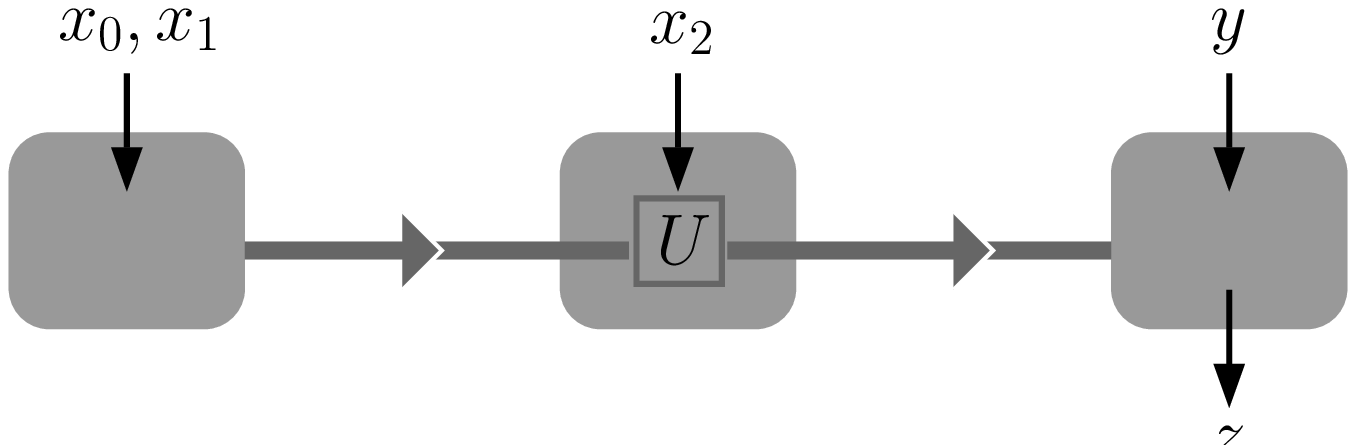}%
}

\caption{$3 \rightarrow 1$ distributed RACs with three different resources are shown. The first two devices receive inputs $x_0,x_1$ and $x_2$ (where $x_i \in \{0,1\}$), respectively. The third device receives input $y \in \{0,1,2\}$ and provides an answer $z$. All classical (double line) or quantum (thick line) communication channels between devices are restricted to two dimensional systems. In (a), $\mathcal{CR}$ denotes that the devices can share classical randomness. In (b), $|\Phi\rangle_i$ represents the $i$-th qubit of the entangled GHZ state shared among the devices. In the distributed EARAC, the first two devices measure the respective shared qubits in two different measurement settings depending on $x_0\oplus x_1$ and $m_1\oplus x_2$, and obtain binary measurement outcome $a$ and $b$, respectively. For different guessing function $f(x,y)$, the explicit forms of these measurements differ, while the third device always chooses the mutually unbiased bases $\sigma_Y,\sigma_X,\sigma_Z$ for $y=0,1,2$. In (c), the first device prepares a qubit according to $x_0,x_1$ and sends it to the second device. For $x_2=1$, the second device applies an unitary $U$ on that qubit, and the third one always measures the qubit in one of the mutually unbiased bases $\sigma_Y,\sigma_X,\sigma_Z$. The encoding quantum states and the unitary operation $U$ are different for different guessing function $f(x,y)$.}
\label{fig}
\end{figure}

The EARAC protocol (see Fig. \ref{fig}) for $f(x,y)=x_y$ is described as follows.
 The three devices, possessed by, say Alice, Bob and Charlie, share multiple copies of the GHZ state $\frac{1}{\sqrt{2}}(|000\rangle + |111\rangle)_{ABC}$. Alice measures her qubit in the basis, $\{\frac{1}{\sqrt{2}}(|0\rangle + e^{- i \phi} |1\rangle), \frac{1}{\sqrt{2}}(|0\rangle - e^{-i \phi}  |1\rangle) \}$ taking $\phi = \frac{\pi}{4}\left(1+2(x_0\oplus x_1)\right)$ and obtains a measurement result $a$. She sends a one bit message $m_1 = a\oplus x_0$ to Bob who proceeds to measure in the basis $\{ \cos\theta |0\rangle + \sin \theta |1\rangle,  \sin \theta |0\rangle - \cos \theta |1\rangle \}$ by choosing $\theta = \cos^{-1}\left( \sqrt{\frac{\sqrt{3}+ (- 1)^{m_1\oplus x_2}}{2\sqrt3}} \right)$. Denoting the measurement outcome as $b$, Bob communicates $m_2 = m_1\oplus b$ to Charlie who measures in the following three bases $\{\ket{\circlearrowright}, \ket{\circlearrowleft}\}, \{\ket{+},\ket{-}\},\{\ket{0},\ket{1}\}$ for inputs $y=0,1,2$ respectively. Charlie's measurement outcome is denoted as $c$ and his guess for $x_y$ will be $m_2\oplus c$. One can obtain,
\begin{widetext}
\begin{equation}\begin{split}\label{ghz}
|\Phi\rangle_{ABC} = & \frac{1}{\sqrt{2}} (|000\rangle + |111\rangle)_{ABC}
=\frac{1}{2} \sum_{a=0,1} (|0\rangle + (-1)^a e^{- i \phi} |1\rangle)_A (|00\rangle +(-1)^a e^{i\phi} |11\rangle)_{BC}, \\&
(|00\rangle +(-1)^a e^{i\phi} |11\rangle)_{BC} = (\cos\theta |0\rangle + \sin \theta |1\rangle)_B (\cos(\theta) |0\rangle +(-1)^a e^{i \phi} \sin(\theta) |1\rangle)_C \\
& \qquad \qquad \qquad \qquad \qquad + (\sin \theta |0\rangle - \cos \theta |1\rangle)_B (\sin(\theta) |0\rangle - (-1)^a e^{i \phi} \cos(\theta) |1\rangle)_C.
\end{split}\end{equation}
\end{widetext}
Following such decomposition, we can check that the distributed EARAC protocol allows us to generate the same quantum states, given in \eqref{qs}, as for the standard EARAC scenario. \\
In fact the above Eq. (\ref{ghz}) is valid for any $\phi$. Thus, other suitable tasks can be constructed by choosing $\phi$ to be $(\phi + \phi')$ for $\phi' = \pi, \frac{\pi}{2}, \frac{3\pi}{4}$, as listed in Table \ref{tab1}, for which the EARAC protocol gives the maximal success probability.

To obtain the maximal value $P^Q$, eight different quantum states corresponding to the vertices of the cube in Bloch sphere should be realized in the third device. It can be noticed from Fig. \ref{fig2} that the four vertices for $x_2=1$ are just the reflection in the $X-Y$ plane [$\mathcal{R}_{XY}$] of the four vertices for $x_2=0$. The EARAC protocol  essentially allows to implement $\mathcal{R}_{XY}$ transformation with respect to $x_2$. The choice of $\phi'$ corresponds to the additional rotation along $Z$ direction [$R_{Z}(\phi')$] followed by $\mathcal{R}_{XY}$. Similarly, one can obtain the other EARACs based on $\mathcal{R}_{XZ}$ and $\mathcal{R}_{YZ}$, in which the measurement bases given in the above EARAC protocol for $x_0\oplus x_1, m_1\oplus x_2$  will be rotated by $R_X(\pi)$ and $R_Y(\pi)$ respectively. Further eight different EARACs corresponding to the different forms of $f(x,y)$ can be constructed with transformations $\mathcal{R}_{XZ},\mathcal{R}_{XZ}R_Y(\phi'),\mathcal{R}_{YZ},\mathcal{R}_{YZ}R_X(\phi')$. \\

\section{Multipartite nonlocality from distributed EARAC}
In this section, we explore the relation between the presented EARAC protocol and quantum nonlocality. Lets consider a three party Bell scenario in which Alice, Bob receive binary inputs $z_1,z_2 \in \{0,1\}$ and Charlie receives inputs $y \in \{0,1,2\}$ and produces a binary output $a,b,c\in \{0,1\}$ respectively. We denote the observed probability of such an event by $P(a,b,c|z_1,z_2,y)$. Following the EARAC protocol, we obtain
\begin{equation}\begin{split}
& x_0\oplus a \oplus b \oplus c = x_y,\\
&\text{where } z_1 = x_0\oplus x_1, z_2 = x_0 \oplus a \oplus x_2.\\
&\implies
a \oplus b \oplus c =
  \begin{cases}
    0, &\text{if } y=0\\
    z_1,&\text{if } y=1\\
    z_2\oplus a,&\text{if } y=2\\
  \end{cases}
\end{split}\end{equation} Thus the distributed EARAC task can be written in the form of the following Bell expression,
\begin{equation}\begin{split}
&B := \frac{1}{12} \sum_{a,b,c,z_1,z_2 \in \{0,1\}} \big[ P(a \oplus b \oplus c = 0|z_1,z_2,0) + \\ & P(a \oplus b \oplus c = z_1|z_1,z_2,1) +P(b \oplus c = z_2|z_1,z_2,2) \big].
\end{split}\end{equation} To probe the nontrivial connection between EARAC and Bell-type inequality, let's assume that the sampling space for Charlie's input is biased. We study two different situations, represented by $t \in \{0,1\}$, where the probability distribution of getting $y=(0,1,2)$ is $(\frac{1}{3}+q,\frac{1}{3}+(-2)^tq,\frac{1}{3}+(-2)^{t \oplus 1}q)$ for some $q \in [0,\frac{1}{6}]$. Subsequently, the Bell expression is modified as,
\begin{equation}\begin{split}
B(t,q) := & \frac{1}{4} \sum_{a,b,c,z_1,z_2 \in \{0,1\}} \left(\frac{1}{3}+q\right) P(a \oplus b \oplus c = 0|z_1,z_2,0) \\ & + \left(\frac{1}{3}+(-2)^{t}q\right) P(a \oplus b \oplus c = z_1|z_1,z_2,1) \\ &+ \left(\frac{1}{3}+(-2)^{t\oplus 1}q\right) P( b \oplus c = z_2|z_1,z_2,2).
\end{split}\end{equation} The maximum value of $B(t,q)$ has been obtained for local (L) and no-signaling bilocal (NSBL) correlations \cite{nsbl,nsbl1}. For the local correlations, we have considered all the possible deterministic strategies. Whereas to get the upper bound for NSBL, we generated all the extremal points (vertices) of bipartite NS polytope. When Alice and Bob share NS correlations i.e. the scenario with two inputs each with binary outputs, all the 24 extremal points are given in \cite{ns22}. In the case where Alice-Charlie or Bob-Charlie share NS correlations, we can obtain all the extremal points (for two and three inputs with binary outputs) by using linear programming. The whole space containing the NS polytope is 24 dimensional. This polytope has 128 vertices, among which 32 are local deterministic and the rest are nonlocal. Using these, the following relation has been observed,
\begin{equation}\begin{split}\label{nsbl}
B(0,q) \overset{L/NSBL_{AB}}{\leq} \frac{2}{3} + \frac{q}{2} \overset{NSBL_{BC}}{\leq} \frac{5}{6} - \frac{q}{2},\\
B(1,q) \overset{L/NSBL_{AB}}{\leq} \frac{2}{3} + \frac{q}{2} \overset{NSBL_{AC}}{\leq} \frac{5}{6} - \frac{q}{2},
\end{split}\end{equation} where $NSBL_{AB}$ denotes the correlation when Alice and Bob share NS resources and so on. For the quantum strategy described earlier, $B(t,q)=P^Q$ independent of $t,q$. It is obtained that sharing convex combinations of $NSBL_{AC}$ and $NSBL_{BC}$ performs better than the EARAC protocol.
However, it can be readily checked from \eqref{nsbl} that if they share NS resources with any particular bipartition, then for $q>\frac{2-\sqrt{3}}{3} $ and the suitable value of $t$, the EARAC protocol over performs. This implies that the EARAC protocol witnesses not only nonlocality but also genuine tripartite nonlocality. \\

\section{Quantum distributed random access code}
In \cite{network}, it is shown that there exists a $3\rightarrow 1$ distributed task where the QRAC leads to the maximal success probability. Let's consider one such QRAC in which $f(x,0)=x_0\oplus x_2 , f(x,1)=x_1 , f(x,2)=x_2$.
The encoding states $|\psi(\theta,\phi)\rangle$ in Eq.\eqref{qs} for the maximum success probability can be prepared by taking
$\theta = \cos^{-1}\left( \sqrt{\frac{\sqrt{3}+ (- 1)^{x_2}}{2\sqrt3}} \right), \phi = \frac{\pi}{4}\left(1+4(x_0\oplus x_2)+2(x_0\oplus x_2 \oplus x_1)\right)$.
The decoding strategy is the same as considered before. The preparation device prepares one of the four states while assuming $x_2=0$ and sends it to transformation device. If $x_2=1$, then the transformation device applies a unitary corresponds to $R_X(\pi)$ ($\pi$ rotation about $X$-axis) in the Bloch sphere before communicating it to measurement device. This protocol leads to the quantum states mentioned above, thereby ensuring that the success probability is maximal.\\

Let us emphasize here the key requirement for the above QRAC strategy. The preparation device sends one of the four different states for the inputs $x_0x_1$ while assuming $x_2=0$, and if $x_2=1$, the transformation device applies some rotation of the Bloch sphere such that the eight states (consisting of initial four states and four states transformed by the unitary) are the eight vertices of a cube in the Bloch sphere. Hence, we need to know all the possible rotations for which four vertices of a regular cube can be transformed exactly to the other four vertices. The symmetric rotation group of a cube contains 24 elements among which one is identity and in 9 cases the rotation axis intersects with the two vertices. If the rotation axis intersects with vertices then those vertices will remain the same after the transformation, and thus eight different states will not be realized. The other suitable 15 rotations are $R_i(\pi/2),R_i(\pi),R_i(3\pi/2)$ for $i \in \{X,Y,Z\}$ and $R_{X\pm Y}(\pi),R_{Y\pm Z}(\pi),R_{Z\pm X}(\pi)$. Depending on these 15 rotations, one can readily construct task such that the QRAC success probability is $P^Q$. The explicit form of $f(x,y)$ will also depend on the initial four vertices of the cube. Few examples of these tasks are given in Table (\ref{tab1}).\\

\section{Comparison between EARAC and QRAC}
First we show that for $f(x,y)=x_y$, the QRAC success probability is strictly less than $P^Q$.\\
{\it Proof-} If the operation of the transformation device on the received qubit is unitary then it is trivial that the corresponding eight quantum states cannot be reproduced. This is because the four states for $x_2=0$ are just the reflection of the states for $x_2=1$, in the $X-Y$ plane of the Bloch sphere. In general, the transformation could be a completely positive trace preserving map upto a unitary. The image of a Bloch sphere of pure states under such a map is an ellipsoid,
\begin{equation}
\begin{split}
\left(\frac{X-t_1}{\lambda_1}\right)^2 +  \left(\frac{Y-t_2}{\lambda_2}\right)^2 +\left(\frac{Z-t_3}{\lambda_3}\right)^2 =1
\end{split}\end{equation} contained within the Bloch sphere. Moreover there is a necessary condition to be completely positive \cite{cp},
\begin{equation} \label{lambdac}
(\lambda_1 + \lambda_2)^2 \leq (1+\lambda_3)^2  - t_3^2.
\end{equation}  Now if the transformation is not a rotation then the only way to have four vertices for $x_2=0$, when the image of Bloch sphere is given by the following ellipsoid,
\begin{equation}
\begin{split}
\left(\frac{X}{\sqrt{2/3}}\right)^2 +  \left(\frac{Y}{\sqrt{2/3}}\right)^2 +\left(\frac{Z-\frac{1}{\sqrt{3}}}{\lambda_3}\right)^2 =1.
\end{split}\end{equation}
From the inequality \eqref{lambdac}, we obtain $\lambda_3 \geq \sqrt{3} -1 \approx 0.732$ but since the ellipsoid is contained within the Bloch sphere, $\lambda_3 \leq 1 - \frac{1}{\sqrt{3}} \approx 0.423$ contradicts that such a transformation exists. If the preparation device prepares these four states assuming $x_2=0$ then the transformation should reproduce the other four states for which the same argument holds. \qed
Similarly, it can be easily shown that for the other EARACs based on $\mathcal{R}_{XZ}, \mathcal{R}_{YZ}$ similar argument holds.
In the Appendix I, the best possible QRAC strategies for some tasks, listed in Table \ref{tab1}, obtained from see-saw [SW] method \cite{seesaw10} are provided.\\
On the other hand, we have checked the upper bounds of EARAC implementing almost quantum [AQ] correlation \cite{aq} in tripartite Bell scenario. Since, the almost quantum set is larger than the quantum set, the upper bounds in Table I, ensure that no EARAC protocol yields the optimal success probability in last four tasks.
\\

\begin{table}[http]
\centering
\begin{tabular}{c| c |c | c}
 $f(x,0),f(x,1),f(x,2)$ & Transformation & {EARAC}&{QRAC} \\
 & in respect to $x_2$ & & \\
\hline
\hline
 $x_0,x_1,x_2$ & $\mathcal{R}_{XY}$ & $P^Q$ & $0.75[SW]$   \\
\hline
$x_0\oplus x_2 , x_1\oplus x_2 , x_2$ &$\mathcal{R}_{XY}R_{Z}(\pi)$ & $P^Q$ & $0.7697[SW]$  \\
\hline
$x_0 \oplus x_2(x_0 \oplus x_1) ,$  &$\mathcal{R}_{XY}R_{Z}(\frac{\pi}{2})$ & $P^Q $ & $0.7546[SW]$ \\
$x_1 \oplus x_2(\overline{x_0 \oplus x_1}) , x_2$  &  &  \\
\hline
$x_0 \oplus x_2(\overline{x_0 \oplus x_1}),$ &$\mathcal{R}_{XY}R_{Z}(\frac{3\pi}{2})$ &  $P^Q $ & $0.7546[SW]$  \\
$x_1 \oplus x_2(x_0 \oplus x_1) , x_2$ &  & \\
\hline
$x_0\oplus x_2   , x_1 , x_2$  & $R_X(\pi)$ &  0.7442[AQ] &  $P^Q$ \\
\hline
$x_0, x_1, x_2\oplus x_0 $   &$R_X(\frac{3\pi}{2})$ &  0.7697[AQ] & $P^Q$ \\
\hline
$x_0 \oplus x_2, x_1 , x_0$   & $R_X(\frac{\pi}{2})$ & 0.7697[AQ] &  $P^Q$ \\
\hline
$x_0 \oplus x_2, x_1 \oplus x_2,  x_0$  & $R_Z(\pi)$ &  0.7697[AQ] & $P^Q$  \\
\hline
\end{tabular}
\caption{List of eight different tasks where EARAC gives the maximal success probability ($P^Q\approx 0.79$) for the first four, whereas QRAC for the rest. The bounds for EARAC is obtained using semi-definite programming implementing {\it almost quantum} [AQ] correlations \cite{aq} in tripartite Bell scenario. The bounds of QRAC are obtained by see-saw method \cite{seesaw10}. The success probability using classical resources is $\frac{2}{3}$ in all the cases.}
\label{tab1}
\end{table}

\section{Experimental Realization}
A proof of principle experimental demonstration of four different $3\rightarrow 1$ distributed QRACs, (five to eight in Table \ref{tab1}), will now be presented.

For the state preparation in our experiment, we have used a heralded single photon source where a 2~mm BBO crystal is pumped using a 390~nm pulsed laser. The pumped photon is down converted into two photons through a process commonly known as spontaneous parametric down-conversion (SPDC). This method is extensively used for the generation of single photons where the idler photon is utilized as a trigger whereas the signal photons are used as single photons. The trigger photon is subsequently detected by a single photon detector $D_T$. The signal photons on the other hand are passed through a narrowband (3~nm) filter before being collected in a single mode fiber (SMF). This ensures that the single photon source is spectrally and spatially very well-defined. We characterized the heralded single photon source and estimated the ratio of multiphoton to single photon pair emission to be below 0.15$\%$. This indicates that our source is a good approximation to a single photon source with negligible higher order contribution. The source provided on average 14.000 photons per second and our single photon coupling efficiency into the SMF is $\sim$20$\%$. The initial polarization state of the photons is prepared in $|H\rangle$ by the use of a fiber polarization controller.
\\
Experimentally, our two level quantum system is realized by using a single photon's polarisation states. The two orthogonal polarization states: $|H\rangle$ and $|V\rangle$ are used for this purpose, where ($H$) and ($V$) are the horizontal and vertical polarization modes of the photon. Information is encoded using the following basis states: $|0\rangle \equiv |H\rangle$, $|1\rangle \equiv |V\rangle$. Therefore, any qubit state can then be represented as $a |0\rangle + b |1\rangle$. The experimental setup is shown in Fig. 3 where Alice prepares any four of her input states, $|\psi_{x_0x_1}\rangle$ with $x_0,x_1\in\lbrace0,1\rbrace$, using a combination of four wave plates and they are parametrized as
\begin{equation}
|\psi_{x_0x_1}\rangle = \cos(2\alpha)|H\rangle+e^{i\phi_{x_0x_1}} \sin(2\alpha)|V\rangle
\label{eq1}
\end{equation}
Suitable orientation of half-wave plate [HWP]($\alpha$) defines the encoding state $|\psi\rangle$ and the combination of the quarter-wave plate [QWP]($\theta_1=45^\circ$),HWP($\beta$),QWP($\theta_2=45^\circ$) ensures the right phase in the encoding state. This robust configuration allows Alice to prepare any of the required states by suitable orientations of the two HWPs. The corresponding wave plate settings for preparing all four of Alice's states are shown in table \ref{tab1exp} for all the four QRAC tasks.

In the case where $x_2=0$, no unitary rotation is performed by Bob. This is equivalent to a scenario where the wave plates of Bob are oriented at QWP($45^\circ$), HWP(-$45^\circ$), QWP($45^\circ$) respectively. The encoding state is then communicated to the measurement device as it is. However for $x_2=1$, Bob performs a unitary rotation about some axis on the Bloch sphere. The unitary rotations considered in the scope of this experiment correspond to the following rotations $R_X(\pi)$, $R_X(\frac{3\pi}{2})$, $R_X(\frac{\pi}{2})$ and $R_Z(\pi)$. For this purpose, a combination of QWP($\theta_3$),HWP($\gamma$) and QWP($\theta_4$) is used. Suitable orientations of QWP($\theta_3$),HWP($\gamma$) and QWP($\theta_4$) allow Bob to perform the specific rotation about $X$-axis or the $R_Z(\pi)$ rotation (Fig. 3). In this proof of principle experimental demonstration, the values of $x_0,x_1,x_2$ were preselected rather than a randomized selection. However, a randomized selection between the settings choice can be implemented by mounting all the corresponding wave plates on motorized stages. The appropriate wave plate settings for implementing the above unitary rotations of Bob are presented in Table \ref{tab1exp}.

\begin{widetext}

\begin{figure}[http]
\begin{center}
\includegraphics[width=11cm,height=3cm]{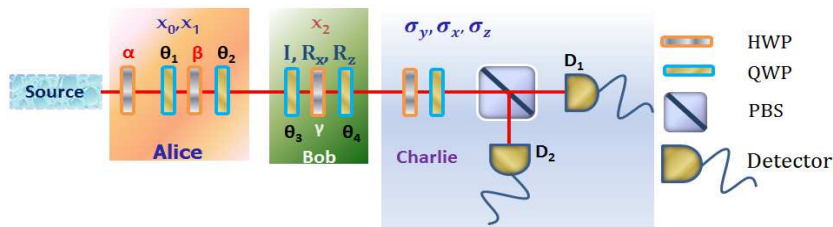}
\caption{Experimental set-up for $3\rightarrow 1$ distributed QRAC. Alice encodes her states in horizontal and vertical single photon polarization states that are prepared by suitable orientation of HWP($\alpha$) and the combination of QWP($\theta_1$),HWP($\beta$),QWP($\theta_2$). Unitary rotations by Bob along x-axis, z-axis and $\eins$ transformation are implemented by a combination of QWP($\theta_3$),HWP($\gamma$) and QWP($\theta_4$) respectively. A combination of HWP, QWP and PBS followed by two single photon detectors $D_i$ ($i=1,2$) allow Charlie to perform the measurements in $\sigma_Y,\sigma_X,\sigma_Z$ bases respectively. }
\label{expsetup}
\end{center}
\end{figure}

\begin{table}[http]
\begin{tabular}{|c| c | c|c| c | c|c| c | c|c|c|}\hline
{QRAC Task} & {State}& \multicolumn{4}{|c|}{Alice's settings} & \multicolumn{5}{|c|}{Unitary by Bob}\\
\hline
 & & & & & &$x_2=0$ &\multicolumn{3}{|c|}{$x_2=1$} & \\ 
$f(x,0),f(x,1),f(x,2)$  & {$\psi_{x_0x_1}$}&$HWP(\alpha)$&$QWP(\theta_1)$&$HWP(\beta)$&$QWP(\theta_2)$ & &$QWP(\theta_3)$&$HWP(\gamma)$&$QWP(\theta_4)$ & \\ 
\hline

 & $\psi_{00}$ &  $13.6839^\circ$ &  $45^\circ$ &  $-56.25^\circ$&  $45^\circ$& $\eins$ &  $90^\circ$ &  $45^\circ$&  $90^\circ$ & \\

  & $\psi_{01}$ &  $13.6839^\circ$ &  $45^\circ$ &  $-78.75^\circ$&  $45^\circ$& $\eins$ &  $90^\circ$ &  $45^\circ$&  $90^\circ$ & \\

$x_0\oplus x_2   , x_1 , x_2$  & $\psi_{11}$ &  $13.6839^\circ$ &  $45^\circ$ &  $-101.25^\circ$&  $45^\circ$& $\eins$ &  $90^\circ$ &  $45^\circ$&  $90^\circ$ & $R_X(\pi)$\\

  & $\psi_{10}$ &  $13.6839^\circ$ &  $45^\circ$ &  $-123.75^\circ$&  $45^\circ$& $\eins$ &  $90^\circ$ &  $45^\circ$&  $90^\circ$ & \\
\hline

 & $\psi_{00}$ &  $13.6839^\circ$ &  $45^\circ$ &  $-56.25^\circ$&  $45^\circ$& $\eins$ &  $90^\circ$ &  $67.5^\circ$&  $90^\circ$ & \\

  & $\psi_{01}$ &  $13.6839^\circ$ &  $45^\circ$ &  $-78.75^\circ$&  $45^\circ$& $\eins$ &  $90^\circ$ &  $67.5^\circ$&  $90^\circ$ & \\

$x_0, x_1, x_2\oplus x_0 $  & $\psi_{11}$ &  $31.3161^\circ$ &  $45^\circ$ &  $-101.25^\circ$&  $45^\circ$& $\eins$ &  $90^\circ$ &  $67.5^\circ$&  $90^\circ$ & $R_X (\frac{3\pi}{2})$\\

  & $\psi_{10}$ &  $31.3161^\circ$ &  $45^\circ$ &  $-123.75^\circ$&  $45^\circ$& $\eins$ &  $90^\circ$ &  $67.5^\circ$&  $90^\circ$ & \\
\hline

 & $\psi_{00}$ &  $13.6839^\circ$ &  $45^\circ$ &  $-56.25^\circ$&  $45^\circ$& $\eins$ &  $90^\circ$ &  $22.5^\circ$&  $90^\circ$ & \\

  & $\psi_{01}$ &  $13.6839^\circ$ &  $45^\circ$ &  $-78.75^\circ$&  $45^\circ$& $\eins$ &  $90^\circ$ &  $22.5^\circ$&  $90^\circ$ & \\

$x_0 \oplus x_2, x_1 , x_0$  & $\psi_{11}$ &  $31.3161^\circ$ &  $45^\circ$ &  $-101.25^\circ$&  $45^\circ$& $\eins$ &  $90^\circ$ &  $22.5^\circ$&  $90^\circ$ & $R_X (\frac{\pi}{2})$\\

  & $\psi_{10}$ &  $31.3161^\circ$ &  $45^\circ$ &  $-123.75^\circ$&  $45^\circ$& $\eins$ &  $90^\circ$ &  $22.5^\circ$&  $90^\circ$ & \\
\hline

 & $\psi_{00}$ &  $13.6839^\circ$ &  $45^\circ$ &  $-56.25^\circ$&  $45^\circ$& $\eins$ &  $45^\circ$ &  $-90^\circ$&  $45^\circ$ & \\

  & $\psi_{01}$ &  $13.6839^\circ$ &  $45^\circ$ &  $-78.75^\circ$&  $45^\circ$& $\eins$ &  $45^\circ$ &  $-90^\circ$&  $45^\circ$ & \\

$x_0 \oplus x_2, x_1 \oplus x_2,  x_0$  & $\psi_{11}$ &  $31.3161^\circ$ &  $45^\circ$ &  $-101.25^\circ$&  $45^\circ$& $\eins$ & $45^\circ$ &  $-90^\circ$&  $45^\circ$ & $R_Z (\pi)$\\

  & $\psi_{10}$ &  $31.3161^\circ$ &  $45^\circ$ &  $-123.75^\circ$&  $45^\circ$& $\eins$ &  $45^\circ$ &  $-90^\circ$&  $45^\circ$ & \\
\hline
\end{tabular}
\caption{$\psi_{x_0x_1}$ are the quantum states in eq. \ref{eq1}. The orientations of HWP($\alpha$) and the combination of QWP($\theta_1$),HWP($\beta$),QWP($\theta_2$) allow to prepare quantum states $\psi_{x_0x_1}$ for the given QRAC task. $x_2=0$ corresponds to an identity operation from Bob (QWP($\theta_3=45^\circ$), HWP($\gamma=-45^\circ$), QWP($\theta_4=45^\circ$)) on the received quantum state whereas the listed orientations and the combination of QWP($\theta_3$),HWP($\gamma$),QWP($\theta_4$) help to implement the unitary rotations ($R_X(\pi), R_X(\frac{3\pi}{2}), R_X(\frac{\pi}{2})$ and $R_Z(\pi)$) corresponding to the four tasks when $x_2=1$. }
\label{tab1exp}
\end{table}
\end{widetext}
For either of the two cases, $x_2\in \{0,1\}$, Charlie's task is to measure the received state in $\sigma_Y,\sigma_X,\sigma_Z$ bases and to recover the desired outcome with a high success probability $P$. For this purpose, the choice to measure in a specific basis is implemented through a combination of one HWP, one QWP and one polarization beam splitter (PBS) followed by two single photon detectors $D_i (i=1,2)$ in each spatial mode of the PBS. For a measurement in $\sigma_Y$, the HWP and QWP settings correspond to $0^\circ,-45^\circ$. For $\sigma_x$, they correspond to $22.5^\circ,0^\circ$ and $0^\circ,0^\circ$ for the $\sigma_Z$ basis. In this way, for any given rotation (no rotation) performed by Bob, the 4 states are measured in $\sigma_Y,\sigma_X,\sigma_Z$ bases and the success probability of the QRAC is then estimated from the number of detection events in the two single photon detectors.

Silicon avalanche photodiodes (APDs) with an effective detection efficiency $\eta_d = 0.55$ were used at the two output ports of the PBS. These detectors have a dark count rate of $R_d \simeq 400$ Hz and a dead  time of 50~ns. A home built coincidence unit was used to record the number of coincidence events between the signal and idler photons. This multi-channel coincidence logic has a detection time window of $1.7$~ns. 
The total measurement time for each experimental setting was $10$~s. For the experimental setup in Fig. 3, considering the coupling losses, detection efficiencies and the registered counts, the overall heralding efficiency is estimated to be $\sim4\%$.

The average success probabilities ($P^Q_{exp}$) for the four constructed QRAC tasks are presented in Table~\ref{tab2}. Appendix II contains the estimated quantum success probabilities (for all states, operations and measurements) corresponding to each task, provided in tables \ref{tabt1exp}, \ref{tabt2exp}, \ref{tabt3exp} and \ref{tabt4exp} respectively. The average quantum success probabilities in table~\ref{tab2} are in good agreement with the predictions of quantum mechanics, namely for an ideal experiment where $P^Q = 0.7887$. The quality of the optical setup depends upon the intrinsic imperfections in the PBS and wave plates. The used PBS has an extinction ratio of 300:1, and the wave plates have a stated retardation precise up to $\lambda/300$. The estimated errors take into account both the Poissonian counting statistics and the systematic errors. For systematic errors, the main contribution is due to the imperfect wave plate settings and the intrinsic imperfections in the PBS and wave plates.

\begin{table}[http]
\centering
\begin{tabular}{|c| c|}\hline
QRAC Task & $P^Q_{exp}$  \\
\hline
$x_0\oplus x_2   , x_1 , x_2$ & 0.790 $\pm$  0.018 \\\hline
$x_0, x_1, x_2\oplus x_0 $  & 0.787 $\pm$  0.018 \\\hline
$x_0 \oplus x_2, x_1 , x_0$ & 0.788 $\pm$  0.0018 \\\hline
$x_0 \oplus x_2, x_1 \oplus x_2,  x_0$ & 0.788 $\pm$  0.017 \\\hline
\end{tabular}
\caption{Average quantum success probabilities, $P^Q_{exp}$, for the four distributed QRAC tasks. The average values are estimated from the measurements provided in tables \ref{tabt1exp}, \ref{tabt2exp}, \ref{tabt3exp} and \ref{tabt4exp} in Appendix II. The estimated errors take into account both the statistical and the systematic errors.}
\label{tab2}
\end{table}

\section{Conclusions} This work provides a remarkably simple scenario where the two different manifestation of quantum resources are complementary to each other. Although the maximal success probability can be reached for all the tasks that we've considered, it is only true for one of the resources. Interestingly, the optimal construction of EARAC and QRAC protocols correspond to the three reflection symmetries and fifteen rotational symmetries of a cube in Bloch sphere. In the future, we plan to generalize the distributed QRAC to systems of higher dimension to see if this property persists. It would also be interesting to investigate more general networks that might be related to some unexplored aspects of nonlocality. For future experiments, ultra bright state-of-the-art electrically or optically driven single photon sources based on color centers in diamond or semiconductor structures as quantum dots can be utilized instead of a heralded single photon source. This together with highly efficient fiber-coupling solutions and by the use of superconducting nanowire single-photon detectors (SSPDs) with high detection efficiency at the target wavelength can lead to practical realization of this and related quantum communication protocols.  \\

\section*{Acknowledgments} The see-saw method was implemented with SDPT3 solver \cite{TTT99} with the support of YALMIP toolbox \cite{YALMIP} within OCTAVE \cite{octave}. This work is supported by the Swedish Research Council (VR), Knut and Alice Wallenberg Foundation, ERC Advanced Grant QOLAPS and the NCN through Grant No. 2014/14/E/ST2/00020 and DS programs of the Faculty of Electronics, Telecommunications and Informatics, Gdan´sk
University of Technology.

{\textbf{Appendix I: QRAC bounds for the first four tasks in Table \ref{tab1}} \\
Here we will provide bounds for some tasks involving QRAC using the see-saw \cite{seesaw10} method with semi-definite programming \cite{SDP} optimization technique.
There are three steps to model the whole protocol. Let $\ket{\psi_{x_0,x_1}}$ be the initial state preparation, $J(\Phi)$ be the general quantum operation in Choi representation, and $M_y$ is the measurement. The see-saw optimization is implemented in three intertwined stages treating one of $M$, $\ket{\psi}$ or $J(\Phi)$ as a variable, and keeping the remaining two constant.
\begin{widetext}
For the first task in Table \ref{tab1}, the maximum value of $P$ is obtained to be 0.75.
Alice prepares state as $\ket{\psi_{00}} = \ket{0}$, $\ket{\psi_{01}} = \ket{+}$, $\ket{\psi_{10}} = \ket{-}$ and $\ket{\psi_{11}} = \ket{1}$.
The channel for $x_2=1$ in Choi representation is,
\be
	\nonumber
	\begin{bmatrix}
		\frac{3}{4} & \frac{1}{4} & \frac{1}{4} + (-1)^{x_2} \times \frac{i}{4} & -\frac{1}{2} \\
		\frac{1}{4} & \frac{1}{4} & 0 & -\frac{1}{4} + (-1)^{x_2} \times \frac{i}{4}\\
		\frac{1}{4} - (-1)^{x_2} \times \frac{i}{4} & 0 & \frac{1}{4} & -\frac{1}{4} \\
		-\frac{1}{2} & -\frac{1}{4} - (-1)^{x_2} \times \frac{i}{4} & -\frac{1}{4} & \frac{3}{4}
	\end{bmatrix}.
\ee \\

The measurement bases are $M_0 = \{\ket{0},\ket{1}\}$, $M_1 = \{\ket{+},\ket{-}\}$ and $M_2 = \ket{\circlearrowleft}, \ket{\circlearrowright}$.

For the second task in Table \ref{tab1}, the obtained maximum value of $P$ is $\frac{1}{12} (7+\sqrt{5}) \approx 0.7697$.
Alice's encoding states are $\ket{\psi_{x_1 x_2}}$: $\ket{\psi_{00}} = \ket{0}$, $\ket{\psi_{01}} = \ket{\psi_{10}} =  \sqrt{\frac{1}{10} \left(5+\sqrt{5}\right)} \ket{0} + \sqrt{\frac{2}{5+\sqrt{5}}} \ket{1} $ and $\ket{\psi_{11}} =  \frac{1}{\sqrt{5}} \ket{0} + \frac{2}{\sqrt{5}} \ket{1} $.
The channel for $x_2=0$ is
\be
	\begin{bmatrix}
	 \alpha _1 & -\sqrt{\alpha _1\left(1-\alpha _1\right)} & \sqrt{\alpha _1\left(1-\alpha _1\right)}i & \alpha _1i \\
	 -\sqrt{\alpha _1\left(1-\alpha _1\right)} & 1-\alpha _1 & -\left(1-\alpha _1\right)i & -\sqrt{\alpha _1\left(1-\alpha _1\right)}i \\
	 -\sqrt{\alpha _1\left(1-\alpha _1\right)}i & \left(1-\alpha _1\right)i & 1-\alpha _1 & \sqrt{\alpha _1\left(1-\alpha _1\right)} \\
	 -\alpha _1i & \sqrt{\alpha _1\left(1-\alpha _1\right)}i & \sqrt{\alpha _1\left(1-\alpha _1\right)} & \alpha _1 \\
	\end{bmatrix},
\ee
and for $x_2=1$ is
\be
	\begin{bmatrix}
	 \alpha _2 & \sqrt{\alpha _2\left(1-\alpha _2\right)} & -\sqrt{\alpha _2\left(1-\alpha _2\right)}i & \alpha _2i \\
	 \sqrt{\alpha _2\left(1-\alpha _2\right)} & 1-\alpha _2 & -\left(1-\alpha _2\right)i & \sqrt{\alpha _2\left(1-\alpha _2\right)}i \\
	 \sqrt{\alpha _2\left(1-\alpha _2\right)}i & \left(1-\alpha _2\right)i & 1-\alpha _2 & -\sqrt{\alpha _2\left(1-\alpha _2\right)} \\
	 -\alpha _2i & -\sqrt{\alpha _2\left(1-\alpha _2\right)}i & -\sqrt{\alpha _2\left(1-\alpha _2\right)} & \alpha _2 \\
	\end{bmatrix},
\ee
where $\alpha _1=0.5+1\left/\sqrt{5}\right.$ and $\alpha _2=0.5-1\left/\sqrt{5}\right.$.

The measurement bases are $M_0=M_1=\{\ket{0},\ket{1}\}$ and $M_3=  \ket{\circlearrowleft}, \ket{\circlearrowright}$.

For the third task in Table \ref{tab1}, the obtained maximum value of $P$ is $\frac{1}{18} \left(9+\sqrt{21}\right) \approx 0.7546$. The fourth task is the same as this upto a rearrangement of the input $y$.
Alice's encoding states are, $\ket{\psi_{00}} = \ket{0}$, $\ket{\psi_{01}} = \sqrt{\frac{25}{63} \frac{4 \sqrt{34}}{63}} \ket{0} + \frac{1}{3} \sqrt{\frac{1}{7} \left(38-4 \sqrt{34}\right)} \ket{1}$, $\ket{\psi_{10}} =  -\sqrt{\frac{2}{3}} \ket{0} + \frac{1}{\sqrt{3}} \ket{1}$ and $\ket{\psi_{11}} = \frac{2}{\sqrt{21}} \ket{0} + \sqrt{\frac{17}{21}} \ket{1}$.
The channel for $x_2=0$ is
\be
	\begin{bmatrix}
	 \alpha _1 & \sqrt{\alpha _1\left(1-\alpha _1\right)} & \sqrt{\alpha _1\left(1-\alpha _1\right)} & -\alpha _1 \\
	 \sqrt{\alpha _1\left(1-\alpha _1\right)} & 1-\alpha _1 & 1-\alpha _1 & -\sqrt{\alpha _1\left(1-\alpha _1\right)} \\
	 \sqrt{\alpha _1\left(1-\alpha _1\right)} & 1-\alpha _1 & 1-\alpha _1 & -\sqrt{\alpha _1\left(1-\alpha _1\right)} \\
	 -\alpha _1 & -\sqrt{\alpha _1\left(1-\alpha _1\right)} & -\sqrt{\alpha _1\left(1-\alpha _1\right)} & \alpha _1 \\
	\end{bmatrix},
\ee
and for $x_2=1$ is
\be
	\begin{bmatrix}
	 \alpha _2 & -\sqrt{\alpha _2\left(1-\alpha _2\right)} & -\sqrt{\alpha _2\left(1-\alpha _2\right)} & -\alpha _2 \\
	 -\sqrt{\alpha _2\left(1-\alpha _2\right)} & 1-\alpha _2 & 1-\alpha _2 & \sqrt{\alpha _2\left(1-\alpha _2\right)} \\
	 -\sqrt{\alpha _2\left(1-\alpha _2\right)} & 1-\alpha _2 & 1-\alpha _2 & \sqrt{\alpha _2\left(1-\alpha _2\right)} \\
	 -\alpha _2 & \sqrt{\alpha _2\left(1-\alpha _2\right)} & \sqrt{\alpha _2\left(1-\alpha _2\right)} & \alpha _2 \\
	\end{bmatrix},
\ee
where $\alpha _1=\frac{1}{2}+\frac{1}{6} \sqrt{\frac{1}{357} \left(937+160 \sqrt{34}\right)}$ and $\alpha _2=\left.\left(357+51\sqrt{21}-4\sqrt{714}\right)\right/714$.

The measurements bases are, $M_0 = \{\ket{0},\ket{1}\}$, $M_1 =\{ 2 \sqrt{\frac{2}{17}} \ket{0} + \frac{3}{\sqrt{17}} \ket{1}, 2 \sqrt{\frac{2}{17}} \ket{1} - \frac{3}{\sqrt{17}} \ket{0}\}$, and $M_2 = \{ \sqrt{\frac{1}{2}+\sqrt{\frac{2}{17}}} \ket{0} + \frac{3}{\sqrt{34+4 \sqrt{34}}} \ket{1}, \sqrt{\frac{1}{2}+\sqrt{\frac{2}{17}}} \ket{1} - \frac{3}{\sqrt{34+4 \sqrt{34}}} \ket{0}\}$.\\
\\
{\textbf{Appendix II: Experimentally estimated probabilities for the four QRAC tasks}} \\
\begin{table}[http]

\begin{tabular}[t]{|c| c | c|c| }\hline
State & \multicolumn{3}{|c|}{Unitary by Bob ($R_X(\pi)$)}\\ \hline
 {$\psi_{x_0x_1}$}&$P^{\sigma_Y}_{exp}$&  $P^{\sigma_X}_{exp}$& $P^{\sigma_Z}_{exp}$\\ \hline

 $\psi_{00}$ & 0.797 $\pm$	0.017 &0.786 $\pm$	0.019&	0.788 $\pm$	0.023

\\\hline

  $\psi_{01}$ & 0.789 $\pm$ 0.017	&	0.790 $\pm$	0.020&	0.787 $\pm$	0.021
 \\\hline

 $\psi_{11}$ & 0.789 $\pm$	0.017 &	0.791 $\pm$	0.017&	0.794 $\pm$	0.023 \\ \hline

 $\psi_{10}$ &0.793 $\pm$	0.017	  &0.795 $\pm$	0.016	&	0.786 $\pm$	0.020 \\ \hline

\end{tabular}
\hfill
\hspace{-5em}
\begin{tabular}[t]{|c| c | c|c| }\hline

State & \multicolumn{3}{|c|}{Unitary by Bob ($\eins$)}\\ \hline
 {$\psi_{x_0x_1}$}&$P^{\sigma_Y}_{exp}$&  $P^{\sigma_X}_{exp}$& $P^{\sigma_Z}_{exp}$\\ 
\hline
  {$\psi_{00}$} & {0.789 $\pm$		0.014} &	{0.788 $\pm$	0.016}&		{0.789 $\pm$		0.020}	
\\\hline

 {$\psi_{01}$} & {0.788 $\pm$		0.014} 	&{0.786 $\pm$		0.017}&	{0.786	$\pm$	0.018}	
 \\\hline

 {$\psi_{11}$} & {0.788 $\pm$	0.014} &	{0.791 $\pm$	0.018}	& {0.790 $\pm$	0.020}  \\ \hline

 {$\psi_{10}$} & {0.800 $\pm$	0.013}		&{0.785 $\pm$	0.019}&	{0.787 $\pm$	0.018}	 \\ \hline
 \end{tabular}

\caption{QRAC task $x_0\oplus x_2   , x_1 , x_2$: $\psi_{x_0x_1}$ are the quantum states in eq. \ref{eq1}. $P^{\sigma_i}_{exp}$, $i \in \lbrace Y,X,Z \rbrace$, represents the experimentally estimated success probabilities for measurements performed in the $\sigma_Y, \sigma_X, \sigma_Z$ bases. Measurement results for both input values of $x_2$  are shown, which are associated with different unitary operations performed by Bob. The estimated errors take into account both the statistical and systematic errors.}
\label{tabt1exp}
\end{table}

\begin{table}[http]
\begin{tabular}[t]{|c| c | c|c| }\hline
State & \multicolumn{3}{|c|}{Unitary by Bob ($R_X({3\pi}/2)$)}\\ \hline
 {$\psi_{x_0x_1}$}&$P^{\sigma_Y}_{exp}$&  $P^{\sigma_X}_{exp}$& $P^{\sigma_Z}_{exp}$\\ \hline


 $\psi_{00}$ & 0.796 $\pm$	0.019&	0.787 $\pm$	0.015	&	 0.790 $\pm$	0.022\\ \hline
 $\psi_{01}$ & 0.791 $\pm$	0.018	&	0.786 $\pm$	0.016&0.782 $\pm$	0.019	\\ \hline

 $\psi_{11}$ &	0.786 $\pm$	0.019&	0.786 $\pm$	0.016&	0.782 $\pm$	0.022\\ \hline

 $\psi_{10}$ &	0.781 $\pm$	0.018&	0.788 $\pm$	0.016&	0.786 $\pm$	0.020 \\ \hline

\end{tabular}
\hfill
\begin{tabular}[t]{|c| c | c|c| }\hline
 {State}& \multicolumn{3}{|c|}{Unitary by Bob ($\eins$)}\\ \hline
{$\psi_{x_0x_1}$}&$P^{\sigma_Y}_{exp}$&  $P^{\sigma_X}_{exp}$& $P^{\sigma_Z}_{exp}$\\ 
\hline

 $\psi_{00}$ & 0.789 $\pm$		0.014 &	0.788 $\pm$	0.016&	0.789 $\pm$		0.021	
\\\hline

 $\psi_{01}$ &0.788 $\pm$		0.014  &0.786 $\pm$		0.016&	0.786	$\pm$	0.017	
 \\\hline

$\psi_{11}$ &  0.783 $\pm$	0.013	&	0.792 $\pm$	0.017&		0.784 $\pm$	0.020 \\\hline

$\psi_{10}$ & 0.789 $\pm$	0.014	 &	0.788 $\pm$	0.016&0.787 $\pm$	0.017 \\\hline
\end{tabular}

\caption{QRAC task $x_0, x_1, x_2\oplus x_0 $: $\psi_{x_0x_1}$ are the quantum states in eq. \ref{eq1}. $P^{\sigma_i}_{exp}$, $i \in \lbrace Y,X,Z \rbrace$, represents the experimentally estimated success probabilities for measurements performed in the $\sigma_Y, \sigma_X, \sigma_Z$ bases. Measurement results for both input values of $x_2$  are shown, which are associated with different unitary operations performed by Bob. The estimated errors take into account both the statistical and the systematic errors.}
\label{tabt2exp}
\end{table}

\begin{table}[http]
\begin{tabular}[t]{|c| c | c|c| }\hline
State & \multicolumn{3}{|c|}{Unitary by Bob ($R_X(\pi/2)$)}\\ \hline
 {$\psi_{x_0x_1}$}&$P^{\sigma_Y}_{exp}$&  $P^{\sigma_X}_{exp}$& $P^{\sigma_Z}_{exp}$\\ \hline

 $\psi_{00}$ & 0.779 $\pm$	0.017 &	0.798 $\pm$	0.017&	0.786 $\pm$	0.019\\ \hline

 $\psi_{01}$ & 0.785 $\pm$	0.018&		0.788 $\pm$	0.017&	0.792 $\pm$	0.021 \\ \hline

 $\psi_{11}$ & 0.784 $\pm$	0.019&	0.789 $\pm$	0.016&	0.784 $\pm$	0.019 \\ \hline

 $\psi_{10}$ & 0.795 $\pm$	0.019&	0.788 $\pm$	0.016&	0.796 $\pm$	0.022  \\ \hline

\end{tabular}
\hfill
\begin{tabular}[t]{|c| c | c|c| }\hline

 {State}& \multicolumn{3}{|c|}{Unitary by Bob ($\eins$)}\\ \hline
{$\psi_{x_0x_1}$}&$P^{\sigma_Y}_{exp}$&  $P^{\sigma_X}_{exp}$& $P^{\sigma_Z}_{exp}$\\ 
\hline

 $\psi_{00}$ & 0.789 $\pm$		0.014 &	0.788 $\pm$	0.016&	0.789 $\pm$		0.021	\\\hline

   $\psi_{01}$ & 0.788 $\pm$		0.014 	&0.786 $\pm$		0.017&	0.786	$\pm$	0.018	
 \\\hline

  $\psi_{11}$  &0.783 $\pm$	0.014	&	0.792 $\pm$	0.016&		0.784 $\pm$	0.021 \\\hline

  $\psi_{10}$ &0.789 $\pm$	0.014&	0.788 $\pm$	0.016&	0.787 $\pm$	0.018 \\\hline
  \end{tabular}

\caption{QRAC task $x_0 \oplus x_2, x_1 , x_0$: $\psi_{x_0x_1}$ are the quantum states in eq. \ref{eq1}. $P^{\sigma_i}_{exp}$, $i \in \lbrace Y,X,Z \rbrace$, represents the experimentally estimated success probabilities for measurements performed in the $\sigma_Y, \sigma_X, \sigma_Z$ bases. Measurement results for both input values of $x_2$  are shown, which are associated with different unitary operations performed by Bob. The estimated errors take into account both the statistical and the systematic errors.}
\label{tabt3exp}
\end{table}

\begin{table}[http]
\begin{tabular}[t]{|c| c | c|c| }\hline
State & \multicolumn{3}{|c|}{Unitary by Bob ($R_Z(\pi)$)}\\ \hline
 {$\psi_{x_0x_1}$}&$P^{\sigma_Y}_{exp}$&  $P^{\sigma_X}_{exp}$& $P^{\sigma_Z}_{exp}$\\ \hline

 $\psi_{00}$ &0.785 $\pm$	0.016		&0.791 $\pm$	0.015	&0.789 $\pm$	0.022	 \\ \hline

 $\psi_{01}$ & 0.795 $\pm$	0.016	&	0.785 $\pm$	0.016&	0.788 $\pm$	0.021   \\ \hline

 $\psi_{11}$ & 0.787 $\pm$	0.016 &	0.789 $\pm$	0.015	&	0.789 $\pm$	0.021   \\ \hline

 $\psi_{10}$ &0.789 $\pm$	0.016 	&	0.787 $\pm$	0.015	&	0.782 $\pm$	0.022  \\ \hline
\end{tabular}
\hfill
\begin{tabular}[t]{|c| c | c|c| c |}\hline

 {State}& \multicolumn{3}{|c|}{Unitary by Bob ($\eins$)}\\ \hline
 {$\psi_{x_0x_1}$}&$P^{\sigma_Y}_{exp}$&  $P^{\sigma_X}_{exp}$& $P^{\sigma_Z}_{exp}$\\ 
\hline

$\psi_{00}$ &0.789 $\pm$		0.014  &	0.788 $\pm$	0.016&	0.789 $\pm$		0.021	\\\hline

 $\psi_{01}$ & 0.788 $\pm$		0.014 &0.786 $\pm$		0.017&	0.786	$\pm$	0.017	
 \\\hline

  $\psi_{11}$ & 0.783 $\pm$	0.014	&	0.792 $\pm$	0.016&	0.784 $\pm$	0.021 \\\hline

   $\psi_{10}$ & 0.789 $\pm$	0.014 &	0.788 $\pm$	0.016&	0.787 $\pm$	0.019 \\\hline
   \end{tabular}
\caption{QRAC task $x_0 \oplus x_2, x_1 \oplus x_2,  x_0$: $\psi_{x_0x_1}$ are the quantum states in eq. \ref{eq1}. $P^{\sigma_i}_{exp}$, $i \in \lbrace Y,X,Z \rbrace$, represents the experimentally estimated success probabilities for measurements performed in the $\sigma_Y, \sigma_X, \sigma_Z$ bases. Measurement results for both input values of $x_2$  are shown, which are associated with different unitary operations performed by Bob. The estimated errors take into account both the statistical and the systematic errors.}
\label{tabt4exp}
\end{table}

\end{widetext}

\end{document}